\documentclass[sigconf,natbib=true,anonymous=false]{acmart}

\usepackage{booktabs} 
\usepackage{upgreek} 
\usepackage{tikz-cd}
\usepackage{tikz}
\usetikzlibrary{shapes,arrows}
\usepackage{hyperref}
\usepackage{graphicx,verbatimbox}
\usepackage{listings}
\usepackage{amsmath}
\usepackage{subcaption}
\usepackage{mathtools}
\usepackage{color}
\usepackage{tabularx,ragged2e}
\usepackage{multirow}
\usepackage{mathtools}
\usepackage[british]{babel}
\usepackage{enumerate}
\usepackage{enumitem}
\setlist{leftmargin=0mm}

\makeatletter
\newcommand\footnoteref[1]{\protected@xdef\@thefnmark{\ref{#1}}\@footnotemark}
\makeatother

\copyrightyear{2022} 
\acmYear{2022} 
\setcopyright{acmlicensed}\acmConference[SIGIR '22]{Proceedings of the 45th International ACM SIGIR Conference on Research and Development in Information Retrieval}{July 11--15, 2021}{Madrid, Spain}
\acmBooktitle{Proceedings of the 45th International ACM SIGIR Conference on Research and Development in Information Retrieval (SIGIR '22, July 11--15, 2022, Madrid, Spain}
\acmPrice{15.00}
\acmDOI{10.1145/3477495.3531884}
\acmISBN{978-1-4503-8037-9/21/07}

\begin{document}
\fancyhead{}
\title{To Interpolate or not to Interpolate: \\ PRF, Dense and Sparse Retrievers}

\author{Hang Li}
\authornote{Both authors contributed equally to the paper.}
\affiliation{%
  \institution{The University of Queensland}
  \city{Brisbane}
  \country{Australia}
}
\email{hang.li@uq.edu.au}

\author{Shuai Wang}
\authornotemark[1]
\affiliation{%
  \institution{The University of Queensland}
  \city{Brisbane}
  \country{Australia}
}
\email{shuai.wang2@uq.edu.au}
\author{Shengyao Zhuang}
\affiliation{%
  \institution{The University of Queensland}
  \city{Brisbane}
  \country{Australia}
}
\email{s.zhuang@uq.edu.au}
\author{Ahmed Mourad}
\affiliation{%
  \institution{The University of Queensland}
  \city{Brisbane}
  \country{Australia}
}
\email{a.mourad@uq.edu.au}
\author{Xueguang Ma}
\affiliation{%
  \institution{University of Waterloo}
  \city{Waterloo}
  \country{Canada}
}
\email{x93ma@uwaterloo.ca}
\author{Jimmy Lin}
\affiliation{%
  \institution{University of Waterloo}
  \city{Waterloo}
  \country{Canada}
}
\email{jimmylin@uwaterloo.ca}
\author{Guido Zuccon}
\affiliation{%
  \institution{The University of Queensland}
  \city{Brisbane}
  \country{Australia}
}
\email{g.zuccon@uq.edu.au}
\settopmatter{authorsperrow=4}

\renewcommand{\shortauthors}{Li, H and Wang, S and et al.}

\begin{abstract}
	
Current pre-trained language model approaches to information retrieval can be broadly divided into two categories: sparse retrievers (to which belong also non-neural approaches such as bag-of-words methods, e.g., BM25) and dense retrievers. Each of these categories appears to capture different characteristics of relevance. Previous work has investigated how relevance signals from sparse retrievers could be combined with those from dense retrievers via interpolation. Such interpolation would generally lead to higher retrieval effectiveness. 

In this paper we consider the problem of combining the relevance signals from sparse and dense retrievers in the context of Pseudo Relevance Feedback (PRF). This context poses two key challenges: (1) When should interpolation occur: before, after, or both before and after the PRF process? (2) Which sparse representation should be considered: a zero-shot bag-of-words model (BM25), or a learnt sparse representation? To answer these questions we perform a thorough empirical evaluation considering an effective and scalable neural PRF approach (Vector-PRF), three effective dense retrievers (ANCE, TCTv2, DistillBERT), and one state-of-the-art learnt sparse retriever (uniCOIL).
The empirical findings from our experiments suggest that, 
regardless of sparse representation and dense retriever, interpolation both before and after PRF achieves the highest effectiveness across most datasets and metrics.

\end{abstract}

%
%



\maketitle

\section{Introduction}
\label{intro}

Traditional unsupervised (bag-of-words -- BOWs) sparse retrieval models, such as BM25, use exact term matching to retrieve relevant results from the collection. Recent studies have shown that these models are more likely to retrieve results that partially match the query, i.e., with low relevance labels~\cite{wang2021bert}. Although unsupervised sparse models often fail to rank the most relevant results at the top, they often offer high recall. Combined with high efficiency, unsupervised bag-of-words sparse retrieval models like BM25 are still widely used within information retrieval pipelines, often as the initial retrieval stage of a more complex setup. 
To further enhance precision and push highly relevant results to the top, transformer-based dense retrievers (short for learned dense representations) strike a good balance between effectiveness and efficiency compared to traditional unsupervised sparse models and transformer-based deep language model re-rankers~\cite{xiong2020approximate,zhan2020repbert,lin2020distilling,lin2021batch,hofstatter2021efficiently,qu2021rocketqa,ren2021rocketqav2}.
Dense retrievers utilise dual BERT-style encoders to encode queries and passages separately~\cite{lin2020pretrained}; this allows the pre-encoding of passages into embeddings at indexing time and their offline storage. During query time, the query embeddings can be efficiently computed "on-the-fly"~\cite{zhuang2021tilde}, and relevance estimations measured with a simple similarity calculation. Thus, it becomes feasible to perform retrieval over the entire collection using deep language models with efficiency comparable to traditional unsupervised sparse models, but with much higher effectiveness. While dense retrievers are very effective at encoding passages characterised by high relevance labels (i.e. highly relevant passages), they are less effective at identifying passages of lower relevance value~\cite{wang2021bert}. On the other hand, learned sparse models~\cite{dai2019context,lin2021few,nogueira2019doc2query,mallia2021learning,formal2021splade,formal2021spladev1,dai2020context,bai2020sparterm,zhuang2021fast}, also strike a good balance between effectiveness and efficiency compared to traditional unsupervised sparse models and transformer-based deep language model re-rankers~\cite{nogueira2019passage}. They use transformer-based language models to learn term weights, and achieve comparable effectiveness to dense retrievers.

Interestingly, recent studies have found that interpolating sparse and dense retrieval~\footnote{Also known as hybrid models.} results can further enhance retrieval effectiveness~\cite{wang2021bert,lin2020distilling,lin2021few}, suggesting that both groups of retrievers tend to retrieve different relevant signals~\cite{lin2021few,lin2020pretrained}. An aspect that is still unclear in this context is what the contribution of PRF is with respect to the interpolation of sparse and dense models: this is the focus of our paper.

In this paper, we are interested in investigating the interpolation of dense and sparse retrieval results within the context of Pseudo-Relevance Feedback (PRF). Specifically, we adapt the interpolation approach on top of a recently proposed PRF method called Vector-PRF (VPRF)~\cite{li2021pseudo} that is designed for dense retrievers. This PRF method conducts the first round of dense retrieval to get the top retrieved passages' dense vectors and then uses these passage vectors to improve the original query's dense representation, which is then used to perform the second round of dense retrieval. In our experiments, we consider combining dense and sparse retrieval interpolation with VPRF in three different settings: \textbf{Before Vector-PRF (Pre-PRF), after Vector-PRF (Post-PRF)} and \textbf{both (Both-PRF)}. For Pre-PRF interpolation, we interpolate the sparse retriever with the dense retriever in the first round of retrieval, then apply Vector-PRF to generate the new query representation, and perform a second retrieval. For Post-PRF interpolation, we keep the two rounds of VPRF retrieval unchanged,  but perform the sparse interpolation to the second round retrieval results. For Both-PRF, we apply interpolation in both the first round of retrieval and on the final results of VPRF. The research questions that we aim to address in this paper are:

\begin{enumerate}[leftmargin=25pt]
	\item[\textbf{RQ1:}] When is it better to do interpolation? \textbf{Before Vector-PRF (Pre-PRF), after Vector-PRF (Post-PRF) or both (Both-PRF)}?
	\item[\textbf{RQ2:}] For sparse retrievers, which representation is more effective, unsupervised (BOWs) or learned? To address this research question, we consider BM25 and uniCOIL~\cite{lin2021few}, respectively.
	
	
\end{enumerate}

\section{Related Work}

There are two lines of research that are related to our work. The first line of research  investigates the integration of PRF with dense retrievers. \citet{li2021pseudo} proposed a simple PRF method called Vector-PRF which adapted the classic Rocchio PRF method~\cite{rocchio1971rocchio} used on bag-of-words representations, to dense retrievers in a zero-shot manner. Vector-PRF has been shown to improve effectiveness, at  additional minimal efficiency expense. We adopt this method in our paper. \citet{wang2021pseudo} proposed a more complex model that uses a clustering technique to model the PRF signals; this is in turn applied to the ColBERT dense retriever~\cite{khattab2020colbert}. However, the improvements achieved by this method come at the cost of efficiency. \citet{yu2021improving}, on the other hand, proposed the ANCE-PRF model that requires the training of a new query encoder based on the original ANCE~\cite{xiong2020approximate} query encoder. ANCE-PRF achieved significant improvements over ANCE. However, due to the input limit of the BERT-style model (512 tokens~\cite{devlin2018bert}), ANCE-PRF is limited in the amount of feedback it can consider: experimentally, $k=5$ is the maximum PRF depth for MS MARCO. 

The second line of research regards the interpolation of sparse retrieval models and deep language models to further boost effectiveness, especially in terms of recall. \citet{wang2021bert} investigated the interpolation of BM25 and dense retrievers. Their findings suggest that dense retrievers are highly effective in encoding strong relevance signals, but they are not as effective when dealing with weak relevance signals. The interpolation of BM25 and dense retrievers is able to make up for each other's weaknesses: this interpolation can significantly improve the effectiveness of dense retrievers. Furthermore, \citet{lin2021batch,lin2021few} and \citet{arabzadeh2021predicting} also investigated different approaches to combine learned sparse retrieval results with dense retrieval results to improve retrieval effectiveness, and significant improvements are recorded from their experiments.
Importantly, according to~\citet{wang2021bert}, dense retrievers are not so good at dealing with weak relevance signals. Therefore, PRF approaches based on dense retrievers might also inherit this limitation. To the best of our knowledge, there is no previous study that has examined the interpolation of sparse and dense retrievers within the PRF framework.

\section{Methods}

Next, we introduce our method for interpolating sparse retriever and dense retrievers in the context of PRF, and in particular of the Vector-PRF approach~\cite{li2021pseudo}. In this paper, we adopt the same interpolation strategy used by~\citet{wang2021bert}: a linear interpolation between sparse retriever scores and dense retriever scores, as shown in equation~\ref{eq:inter}.

\begin{equation}
	\label{eq:inter}
	s(p) = \lambda\hat{s}_{Sparse}(p) + (1 - \lambda)s_{Dense}(p)
\end{equation}

\noindent that is, the score of a passage is the linear interpolation of the sparse retriever score of the passage and the score of the same passage from the dense model, modulated by a parameter $\lambda$, which controls the contribution of the sparse retriever score to the final score of the passage. In our experiments, \texttt{Sparse} refers to BM25 or uniCOIL, \texttt{Dense} refers to any dense retrieval model we use. This interpolation mechanism is a simple yet effective approach to ``help'' the dense retrievers capture the passages' weak relevance signals.

As mentioned, for PRF, we used the Rocchio Vector-PRF approach of \citet{li2021pseudo}.
The way we do interpolation with PRF can be categorized into three different types; we discuss each type with detail in the following subsections.

\subsection{Pre-PRF Interpolation}

The PRF process often involves two rounds of retrieval~\cite{li2021pseudo}. The first retrieval round is to generate the initial results for preparing PRF feedback candidates. After getting the initial retrieval results, top-$k$ passages for each query from these results are selected as PRF feedback passages and are used to modify the original query representations. Therefore, the interpolation can be performed at either round of retrieval.

For Pre-PRF zero-shot interpolation, we perform the sparse retriever interpolation with the first round of dense retrieval results, then we apply PRF with the interpolated results to generate the new query representations for the second round of retrieval. After the interpolation, the ranking of the passages in the results are likely to be different, affecting the PRF's performance. 

\subsection{Post-PRF Interpolation}

Other than applying interpolation before the PRF to the initially retrieved results, we also apply interpolation to the results after the second round of retrieval with PRF query representations. In this approach, the initial retrieval results are directly used for generating PRF query representations, then the PRF queries are used to perform a second round of retrieval. After the second round of retrieval, the results are then interpolated with the sparse retriever's results to obtain the final results list. 

\subsection{Both-PRF interpolation}
Finally, Both-PRF performs interpolation before and after PRF. To perform Both-PRF, we firstly interpolate the sparse retriever's results with the dense retriever's results, then we apply PRF with the interpolated results to generate the new query representations for second round of retrieval. Then results from the second round of retrieval are again interpolated with the sparse retriever's results to generate the final results list.

\section{Experimental Setup}

To investigate the interpolation of sparse retrievers with dense retriever PRF approaches, we devise a number of empirical experiments aimed at investigating: 1) the impact of interpolation on different dense retriever PRF approaches; 2) the impact of interpolating sparse retrievers before/after/both the PRF; 3) the impact of interpolating on different sparse retrievers, unsupervised (BOWs) or learned.

\textbf{Datasets.} For all of our experiments, we use the TREC Deep Learning Track passage retrieval task 2019~\cite{craswell2020overview} (DL19) and 2020~\cite{craswell2021overview} (DL20). DL19 contains 43 judged queries, while DL20 contains 54 judged queries. The relevance judgement levels for both datasets  range from 0 (not relevant) to 3 (highly relevant). 
We treat passages with relevance label 1 as not relevant when we compute the binary relevance metrics (i.e., MAP, Recall). The passage collection in our experiments is the MS MARCO Passage Ranking Dataset~\cite{nguyen2016ms}, which is a benchmark English dataset for ad-hoc retrieval that contains $\approx$8.8 million passages. The average judgements per query for DL19 and DL20 are 215.3 and 210.9, whereas the MS MARCO Passage Ranking Dataset only has $\approx$1 judgement per query.



\textbf{Baselines.} We include:
\begin{itemize}[leftmargin=*]
	\item \texttt{ANCE}: First stage dense retriever~\cite{xiong2020approximate}. We use the model implemented in Pyserini\footnote{\url{https://github.com/castorini/pyserini/blob/master/docs/experiments-ance.md}}~\cite{lin2021pyserini} for inference;
	\item \texttt{Vector-PRF (VPRF)}: A simple Rocchio PRF approach based on dense retrievers~\cite{li2021pseudo}. We use the model implemented in Pyserini\footnote{\label{vprf}\url{https://github.com/castorini/pyserini/blob/master/docs/experiments-vector-prf.md}}~\cite{lin2021pyserini};
	\item \texttt{TCT ColBERT V2 HN+ (TCTv2)}: A BERT-style distilled dense retriever learned from ColBERT with reduced query/passage embedding dimensions~\cite{lin2021batch}; 
	\item \texttt{TCT ColBERT V2 HN+ VPRF (TCTv2+VPRF)}: The application of the Rocchio VPRF from~\citet{li2021pseudo} on top of TCT ColBERT V2 HN+ dense retriever. This model is also made available by the authors in Pyserini\footnoteref{vprf}~\cite{lin2021pyserini};
	\item \texttt{DistilBERT KD TASB (DBB)}: A DistilBERT-style dense retriever with balanced topic aware sampling training strategy~\cite{hofstatter2021efficiently}. We use the model implemented in Pyserini\footnote{\url{https://github.com/castorini/pyserini/blob/master/docs/experiments-distilbert_tasb.md}}~\cite{lin2021pyserini} by the original authors;
	\item \texttt{DistilBERT KD TASB + VPRF (DBB+VPRF)}: The application of the Rocchio VPRF from~\citet{li2021pseudo} on top of DistilBERT KD TASB dense retriever. This model is implemented by~\citet{li2021pseudo} and available to use in Pyserini\footnoteref{vprf}~\cite{lin2021pyserini}.
\end{itemize}

In our experiments, we use the parameters $\alpha=0.4$, $\beta=0.6$, and PRF depth = 3 for Rocchio VPRF, following the settings recommended by~\citet{li2021pseudo}. In terms of the interpolation parameter $\lambda$, we use $\lambda=0.5$ for all  experiments.  For generating the BM25 runs to be used for interpolation, we use the BM25 implementation provided by Pyserini~\cite{lin2021pyserini} and we use the default parameter values for $k_1$ and $b$ within Pyserini. For generating uniCOIL runs, we also use the pre-built uniCOIL index provided by Pyserini.

\textbf{Evaluation Measures.} We use the official evaluation metrics from DL19 and DL20: nDCG@10 and Recall@1000. We also report MAP as a complementary metric.



\section{Results}

Next, we examine the results of our empirical investigation; for this we follow the research questions we put forward in Section~\ref{intro}. The main results are presented in Table~\ref{tab:rq12}. We note that all the models used in our experiments are provided by the original authors and we do not train a new model. The training of a new model may lead to variations in the effectiveness results because of e.g., the stochastic nature of the weight initialisation of the models. 
However, we observe that~\citet{li2022improving} have shown that the differences observed empirically in terms of model effectiveness across re-training of the same model are minor and not statistically significant. The use of the original model checkpoint for each model then appears a fair choice in the context of our work, and the variability that the retraining of these models would cause seems a non-critical direction to investigate in this short paper. 

\subsection{RQ1: When is it better to do interpolation? Pre-PRF, Post-PRF or Both-PRF?}

\begin{table*}[]
	\resizebox{7in}{!}{

		\begin{tabular}{ccc|ccc|ccc}
			\toprule
			&\textbf{Dataset} &               &               \multicolumn{3}{c|}{\textbf{DL19}}               &                              \multicolumn{3}{c}{\textbf{DL20}}                               \\ \midrule
			\textbf{Sparse Model} &\textbf{Dense Model} &\textbf{PRF-Interpolation}                  &     \textbf{MAP}     & \textbf{nDCG@10} &   \textbf{Recall@1000}    &         \textbf{MAP}          &   \textbf{nDCG@10}   &            \textbf{Recall@1000}            \\
			\midrule

			&ANCE&                   &        0.3710        &      0.6452      &        0.7554        &            0.4076             &        0.6458        &                0.7764                 \\
			&ANCE-VPRF&              &        0.3831        &      0.6512      &        0.7611        &            0.4118             &        0.6479        &                0.7800                 \\	\midrule
			\multirow{4}{*}{{BM25}}&\multirow{4}{*}{{ANCE}}&No-PRF               &        0.4264        &      0.6888      &   \textbf{0.8607}    &            0.4067             &        0.6264        &                0.8643                 \\
			&&Pre-PRF         &        0.3868        &      0.6620      & 0.7638${^\barwedge}$ &        \textbf{0.4175}        &   \textbf{0.6548}    &         0.7911${^\barwedge}$          \\
			&&Post-PRF         & 0.4322${^\barwedge}$ &      0.6885      &   0.8604${^\star}$   &            0.4140             &        0.6353        & \textbf{0.8666}${^{\barwedge\star}}$  \\
			&&Both-PRF  &   \textbf{0.4345}${^\barwedge}$    & \textbf{0.6895}  &   \textbf{0.8607}    &            0.4100${^\barwedge}$            &        0.6274        &                0.8662${^\barwedge}$                \\

			\midrule
			\multirow{4}{*}{{uniCOIL}}&\multirow{4}{*}{{ANCE}}&No-PRF&0.4587 &0.6908   &   0.8459 &0.4644&0.6984 &   0.8482    \\

			&&Pre-PRF   &0.3857${^\barwedge}$ & 0.6602${^\barwedge}$   & 0.7584${^\barwedge}$ &  0.4154${^\barwedge}$& 0.6545${^\barwedge}$ &0.7892${^\barwedge}$         \\
			&&Post-PRF   & 0.4617${^\star}$ &  0.6910${^\star}$   &    \textbf{0.8495}${^\star}$   &    0.4693${^{\barwedge\star}}$  & \textbf{0.7024}${^\star}$  & \textbf{0.8500}${^\star}$  \\
			&&Both-PRF  &   \textbf{0.4622} & \textbf{0.6917}  &   0.8458 &    \textbf{0.4699}${^\barwedge}$ & 0.7012 &   \textbf{0.8500}${^\barwedge}$ \\
			\midrule\midrule
			&TCTv2&                   &        0.4469        &      0.7204      &        0.8261        &            0.4754             &        0.6882        &                0.8429                 \\
			&TCTv2-VPRF&               &        0.4626        &      0.7219      &        0.8377        &            0.4863             &        0.6952        &                0.8462                 \\\midrule
			\multirow{4}{*}{{BM25}}&\multirow{4}{*}{{TCTv2}}&No-PRF              &        0.4474        &      0.7067      &        0.8753        &            0.4468             &        0.6696        &                0.8872                 \\
			&&Pre-PRF        &   \textbf{0.4698}    & \textbf{0.7268}  &        0.8436        & \textbf{0.4879}${^\barwedge}$ &  \textbf{0.6987}   &         0.8455${^\barwedge}$          \\
			&&Post-PRF        & 0.4547${^\barwedge}$ &      0.7045      &        0.8788        &  0.4511${^{\barwedge\star}}$  &        0.6683        & \textbf{0.8918}${^{\barwedge\star}}$ \\
			&&Both-PRF &        0.4574${^\barwedge}$        &      0.7061      &   \textbf{0.8820}${^\barwedge}$   &            0.4490             &        0.6659        &                0.8902${^\barwedge}$                \\\midrule
			\multirow{4}{*}{{uniCOIL}}&\multirow{4}{*}{{TCTv2}}&No-PRF  & 0.4771&0.7245&0.8557&0.4895&0.718&0.8683    \\
			&&Pre-PRF  & 0.4659&0.7246&0.8415&0.4897&0.7083&0.8484${^\barwedge}$         \\
			&&Post-PRF  & 0.4826${^{\barwedge\star}}$& \textbf{0.7347} & 0.8663${^\barwedge}$& \textbf{0.4926} & 0.7184&0.8718${^\star}$\\
			&&Both-PRF & \textbf{0.4831} & 0.7268 & \textbf{0.8606}${^\barwedge}$ & 0.4920${^\barwedge}$ & \textbf{0.7190} & \textbf{0.8723}${^\barwedge}$ \\
			\midrule\midrule
			&DBB&                     &        0.4590        &      0.7210      &        0.8406        &            0.4698             &        0.6854        &                0.8727                 \\
			&DBB-VPRF&                 &        0.4667        &      0.7285      &        0.8479        &            0.4804             &   \textbf{0.7027}    &                0.8767                 \\ \midrule
			\multirow{4}{*}{{BM25}}&\multirow{4}{*}{{DBB}}&No-PRF                &        0.4584        &      0.6993      &        0.8622        &            0.4417             &        0.6491        &                0.8948                 \\
			&&Pre-PRF          &   \textbf{0.4711}    & \textbf{0.7319}  &        0.8526        & \textbf{0.4812}${^\barwedge}$ & 0.6968${^\barwedge}$ &                0.8755                 \\
			&&Post-PRF          & 0.4652${^\barwedge}$ &      0.7053      &        0.8711        &       0.4509${^\star}$        &   0.6522${^\star}$   &                0.8974                 \\
			&&Both-PRF   &        0.4665${^\barwedge}$        &      0.7019      &   \textbf{0.8720}    &            0.4442             &        0.6474        &            \textbf{0.8977}            \\  \midrule

			\multirow{4}{*}{{uniCOIL}}&\multirow{4}{*}{{DBB}}& No-PRF&0.4779 & 0.7288 & 0.8542 & 0.4859 & 0.7041 & 0.8808    \\
			&&Pre-PRF &   0.4726&0.7255&0.8468&0.4815&0.7002&0.8760   \\
			&&Post-PRF& 0.4822 & 0.7298 & 0.8658 & \textbf{0.4915} & \textbf{0.7097} & \textbf{0.8836}   \\
			&&Both-PRF   & \textbf{0.4858}${^\barwedge}$ & \textbf{0.7334} & \textbf{0.8669} & 0.4880${^\barwedge}$ & 0.7062 & 0.8834 \\\midrule
			\bottomrule
		\end{tabular}
	}
\vspace{6pt}
	\caption{The results of all baseline runs and No-PRF, Pre-PRF, Post-PRF and Both-PRF interpolation runs of all models with the Rocchio Vector PRF approach proposed by~\citet{li2021pseudo}. Statistical significance tests are conducted between Pre- and Post-PRF models, significant difference are marked with $\star$. We also tested the statistical significance with Pre-PRF interpolation versus No-PRF interpolation, and Post-PRF interpolation versus No-PRF interpolation, and Both-PRF interpolation versus no-PRF interpolation, significant difference are marked with $\barwedge$. Best performance among each base sparse model is marked as Bold.}
	\label{tab:rq12}
	\vspace*{-15pt}
\end{table*}


To answer the first research question, we perform Pre-PRF interpolation, Post-PRF interpolation, and Both-PRF interpolation. We do this across two sparse retrievers (one neural, uniCOIL, and one not, BM25), and three dense retrivers (ANCE, TCTV2 and DistillBERT).

We first discuss the results among each dense retriever with only PRF or only interpolation involved. In this case, the interpolation with uniCOIL always gives the highest effectiveness compared to using PRF or using the dense retriever along: this is regardless of the dense retriever of choice and dataset. However, when BM25 is used for interpolation, the interpolated results are lower than those achieved by PRF for MAP and nDCG (but not recall@1000), regardless of dense retriever and dataset.

We now consider interpolation with BM25.
Unlike the findings obtained by comparing the use of the dense retriever with either only interpolation or only PRF, better effectiveness is achieved when both PRF and interpolation with BM25 are used. 
The effectiveness of the Post-PRF interpolation condition is very close to that of Both-PRF interpolation; however, Both-PRF interpolation achieves the highest effectiveness most of the times and across all dense retrievers and datasets.
However, the difference between Post-PRF and Both-PRF interpolation is not significant, and Both-PRF interpolation requires some extra computations compared to Post-PRF.
Pre-PRF interpolation tends to enhance early precision, while Post-PRF interpolation tends to enhance deep recall. Overall, however, Pre-PRF interpolation is the least effective of the three interpolation conditions, regardless of dense retriever, metric and dataset.


We now consider interpolation with uniCOIL. Here also we observe that PRF with uniCOIL interpolation can consistently achieve higher effectiveness than using the dense retrievers or sparse retrievers alone.
Furthermore, we also found that using Post-PRF interpolation or Both-PRF interpolation  outperforms results obtained with interpolation but not PRF, regardless of dense retriever and dataset. However, if Pre-PRF interpolation is used, then effectiveness is often similar or lower than when interpolation is used but no PRF. Indeed, Both-PRF can achieve the highest effectiveness most of the times.


Overall, the condition with interpolation both before and after PRF (Both-PRF) showcased often high effectiveness regardless of dense and sparse retriever. The condition with interpolation performed after PRF (Post-PRF) showcased high effectiveness only with the uniCOIL sparse retriever, but not with BM25.


\vspace{-10pt}
\subsection{RQ2: Which sparse retriever is more effective, unsupervised (BOWs) or learned?}

In our experiments we considered BM25 and uniCOIL as sparse retriever. The uniCOIL method is a typical neural sparse retriever trained with contrastive loss that uses BERT to predict impact scores for both query tokens and document tokens. Before performing predictions, uniCOIL uses docTquery-T5~\cite{nogueira2019doc2query} to expand all passages in the collection by adding potential relevant tokens that are not in the original passages. BM25 instead is a traditional BOWs sparse retriever, where the representation is not learnt. We now consider which of these sparse representation is best to combine with the signal from the dense retrievers; we do this bot in the context of PRF (Pre-PRF, Post-PRF, and Both-PRF interpolation) and when not considering PRF (No-PRF).


The results show that the use of uniCOIL guarantees an increase in MAP and nDCG@10 compared to BM25, but a lower recall@1000. This result is valid across all dense retrievers, datasets and for No-PRF, Post-PRF and Both-PRF interpolation conditions.


For the Pre-PRF interpolation method, however, no general trend is found when comparing the two sparse models.
On DL2019, uniCOIL consistently underperforms BM25 in the pre-PRF interpolation condition, with the only exception of MAP when DBB is used as dense retriever.
On DL2020, uniCOIL always outperforms BM25 when the dense retrievers are TCTV2 or DBB, but underperforms BM25 when ANCE is used as dense retriever.


Overall, we found that when performing PRF and interpolating with the Post-PRF and Both-PRF conditions, an unsupervised BOWs sparse retriever leads generally to high recall, while a neural, trained sparse retriever achieves higher MAP and nDCG@10. However, the Pre-PRF condition shows no stable trends in terms of which sparse retriever to use.

\vspace{-5pt}
\section{Conclusion}
Previous work has argued that sparse and dense retrievers encode different relevance characteristics of a document~\cite{lin2020pretrained,wang2021bert}. Because of this, methods for the combination of these two signals have emerged; the simplest method being the interpolation of the scores originating from a sparse and a dense retriever~\cite{wang2021bert}.

In this paper, we conducted an extensive investigation to study the effect of interpolation between sparse and dense retrievers in the context of Pseudo Relevance Feedback for dense retrievers, and in particular for the scalable and effective Vector-PRF method~\cite{li2021pseudo}. 
With this respect, we studied applying BM25 and uniCOIL as sparse retrievers, along with three dense retrievers: ANCE, TCTV2 and DistillBERT. In terms of when to interpolate sparse and dense signals, we considered doing this before PRF, after PRF, and both before and after PRF.


The empirical results show that interpolation often can boost retrieval effectiveness, regardless of the choice of sparse and dense retrievers. Among the choices of when to interpolate, we found that interpolating both before and after the PRF process is the condition that most often lead to substantial gains.


\section*{Acknowledgements.}
Hang Li is funded by the Grain Research and Development Corporation (GRDC), project AgAsk (UOQ2003-009RTX). Shuai Wang is supported by a UQ Earmarked PhD Scholarship and this research is funded by the Australian Research Council Discovery Projects program ARC DP210104043.

\bibliographystyle{ACM-Reference-Format}
\bibliography{references.bib}

\end{document}